\documentclass{emulateapj}
\graphicspath{{figs/}} %
\usepackage[colorlinks,linkcolor=blue,anchorcolor=blue,citecolor=blue]{hyperref}
\usepackage[hyperref]{backref}

\usepackage{graphicx} 
\usepackage{txfonts}
\usepackage{natbib}
\begin{document}

\title{Local Group Analogs in $\Lambda$CDM cosmological simulations}

\author{Meng Zhai\altaffilmark{1,2}, Qi Guo\altaffilmark{3,4}, Gang Zhao\altaffilmark{1,4}, Qing Gu\altaffilmark{3,4}, Ang Liu\altaffilmark{5}}

\altaffiltext{1}{Key Laboratory of Optical Astronomy, National Astronomical Observatories, Chinese Academy of Sciences, Beijing 100012, Peopleʼs Republic of China; gzhao@nao.cas.cn}
\altaffiltext{2}{Chinese Academy of Sciences South America Center for Astronomy, Chinese Academy of Sciences, Beijing 100012, Peopleʼs Republic of China}
\altaffiltext{3}{Key Laboratory for Computational Astrophysics, National Astronomical Observatories, National Astronomical Observatories, Chinese Academy of Sciences, Beijing 100012, Peopleʼs Republic of China; guoqi@nao.cas.cn}
\altaffiltext{4}{University of Chinese Academy of Sciences, No. 19 A Yuquan Road, Beijing 100049, Peopleʼs Republic of China}
\altaffiltext{5}{INAF Osservatorio Astrofisico di Arcetri, Largo E. Fermi, I-50125 Firenze, Italy}

\begin{abstract}

We use semi-analytic galaxy catalogs based on two high-resolution cosmological $\emph{N}$-body simulations, Millennium-WMAP7 and Millennium-II, to investigate the formation of the Local Group (LG) analogs. Unlike previous studies, we use the observed stellar masses to select the LG member (Milky Way (MW) and M31) analogs, and then impose constrains using the observed separation, isolation, and kinematics of the two main member galaxies. By comparing radial and low-ellipticity orbits between the MW and M31, we find higher tangential velocity results in higher total mass, which are 4.4$^{+2.4}_{-1.5}\times$10$^{12}\rm M_{\odot}$ and 6.6$^{+2.7}_{-1.5}\times$10$^{12}\rm M_{\odot}$ for radial and low-ellipticity orbits. The orbits also influence the individual mass distribution of MW and M31 analogs. For radial orbits, the typical host halo masses of the MW and M31 are 1.5$^{+1.4}_{-0.7}\times$10$^{12}\rm M_{\odot}$ and 2.5$^{+1.3}_{-1.1}\times$10$^{12}\rm M_{\odot}$;  for low-ellipticity orbits, the masses are 2.5$^{+2.2}_{-1.4}\times$10$^{12}\rm M_{\odot}$ and 3.8$^{+2.8}_{-1.8}\times$10$^{12} \rm M_{\odot}$. The LG is located primarily in filaments with tails extending toward higher densities up to $\delta\sim4.5$. The dark matter velocity anisotropy parameters $\beta$ of both the MW and M31 analogs are close to zero in the center, increasing to 0.2--0.3 at 50--80 kpc and decreasing slowly outward. The slope is much flatter than computed from the MW satellites, and the amplitude is smaller than traced by halo stars. Values of $\beta$ from different tracers agree at $\sim$120 kpc where $\beta \sim$ 0.2. We also find that model predictions agree broadly with observations in the radial distribution and luminosity function of satellites around the MW and M31.

\end{abstract}


\section{Introduction}

Numerous large-scale observations, including the cosmic microwave background radiation \citep{Dunkley2009,Planck2014}, power spectrum of galaxy clustering \citep{Percival2010}, and acceleration of cosmic expansion \citep{Riess1998,Perlmutter1999}, have elevated the $\Lambda$CDM to the status of a standard cosmology model. However, evidence on galactic and sub-galactic scales is still limited and can mostly be obtained only by a statistical approach \citep[e.g.][]{Umetsu2015}.

The Local Group (LG) is the closest galaxy group, and consists of the Milky Way (MW), M31 and their satellite galaxies. Their close proximity provides the possibility of detailed comparison between their observed properties and predictions from the concordance cosmology. Well-known discrepancies include the `missing satellite' problem, where orders of magnitude more substructures are identified in the $\Lambda$CDM simulations than the number of observed satellite galaxies \citep{Kauffmann1993,Moore1999,Klypin1999,Mutlu-Pakdil2018,Newton2018},  the `too-big-to-fail' problem, that is, the most luminous observed satellites do not seem to live in the most massive dark subhalos expected from cosmological simulations, which however are `too big to fail' at forming stars \citep{Strigari2010,Boylan2011}, the `core-cusp' problem, where halos are less cuspy than the $\Lambda$CDM predictions \citep{Simon2005, Navarro2010}, and the `planes of satellites' problem, where the observed satellite galaxies are distributed on a much thinner plane \citep{Lynden-Bell1982,Metz2009,Pawlowski2012,Ibata2013} than substructures predicted by the $\Lambda$CDM model.
Some of these discrepancies can be alleviated by reducing the expected host halo mass, for example, having fewer (massive) substructures in less-massive halos so that both the `missing-satellite' problem and the `too-big-to-fail' problem could be less severe  \citep{Wang2012,Cautun2014}. It is thus important to quantify the masses of the MW, M31 and the LG.

One of the most simple and popular ways to determine the mass of the LG halo was firstly proposed by \citet{Kahn1959}, which treats the MW and M31 as point sources approaching each other on a radial orbit (timing augments). They separate from each other after the big bang along with the expansion of the universe, and then decelerate under the gravity until they start to approach each other. A single apocentric passage gives a lower limit of the total mass as $\sim 5\times 10^{12} \rm M_{\odot}$. Calibrations taking into account the structure and the dynamical evolution of the systems were then performed by \citet{Kroeker1991} and \citet{Li2008}, who found the standard timing estimate is an almost unbiased estimate of the viral mass of the LG.  The number was then reduced somewhat due to the updated lower radial velocity, but is still much higher than the sum of the individually estimated mass of the MW and M31 \citep{vanderarel2012}.

Thanks to the advancements in numerical techniques and supercomputers, it is possible to trace the formation and evolution of galaxy populations throughout orders of magnitude, from cD galaxies all the way to the faintest galaxies observed so far, in a single cosmological simulation.
Recent works \citep{Gonz2014,Fattahi2016,Carlesi2017} select halo pairs as LG analogs from $\emph{N}$-body cosmological simulations, which mimic similar host halo masses, separation and kinematics as the MW and the M31. Detailed analysis reveals that the tangential velocity between M31 and the MW plays an important role, varying the final LG mass by a factor of $\sim$ 2. For example, \citet{Carlesi2017} found an LG mass of $\sim1.8\times 10^{12}{\rm M_{\odot}}$ adopting the lower tangential velocity of 17 km $\rm s^{-1}$ \citep{Sohn2012}, while \citet{Carlesi2017} and \citet{McLeod2017} found a much higher LG mass, $\sim3.5\times 10^{12}{\rm M_{\odot}}$, adopting a the higher tangential velocity of 164 km $\rm s^{-1}$.

In the previous simulations, halo mass has usually been used to select dark matter halos in simulations \citep[e.g.][]{Sawala2015,Fattahi2016}. However, while luminosity is a property that can be observed directly, the relation between luminosity (and the inferred stellar mass) and halo mass has been found to show a large scatter \citep{Guo2010, Moster2011,Mandelbaum2006}.
In particular, \citet{Guo2015} found that the probability that a halo with given mass hosts a MW analog is low, with a maximum of $\sim$20 \% at $\rm M_{halo}=$10$^{12} \rm M_{\odot}$. Therefore, in this work we use stellar mass, which can be derived from the observed luminosity and colors, to identify the MW and M31 analogs. We then study the properties of the LG analogs, their host halos, and the satellite galaxies.

This paper is organized as follows. In Section 2, we introduce the selection methods of our galaxy sample, including the two $\emph{N}$-body simulations and the selection criteria for LG candidates. Section 3 presents our results of the LG analogs and their satellite galaxies. A discussion and conclusions are given in Section 4 and Section 5 respectively.

\section{Sample selection}
\subsection{The simulations}
We use two semi-analytical galaxy catalogs from \citet{Guo2013}, which are constructed from two $\emph{N}$-body cosmological simulations, MR7 and MRIIscWMAP7. Both simulations assume $Wilkinson Microwave Anisotropy Probe$ ($WMAP$)-7 yr
(seven-year results from the $WMAP$) cosmological parameters in the standard $\Lambda$CDM model with $\Omega_{\rm b}$=0.045,
$\Omega_{\rm M}$=0.272, $\Omega_{\Lambda}$=0.728 and  $h$ = 0.704. MR7 is the WMAP-7 yr cosmology version of the Millennium Simulation \citep{Springel2005} with a side length of 500~Mpc/$h$. It follows 2160$^{3}$ dark matter particles from redshift 127 to the present day, with a particle mass of $\rm m_{p}$=9.3639$\times$10$^{8}$$\rm M_{\odot}/$$h$. The Plummer-equivalent force softening is $\varepsilon$=5~kpc$/h$. The large volume of MR7 enables us to study the LG analogs in a statistical way.

MRIIscWMAP7 is generated from the Millennium II Simulation \citep{Boylan-Kolchin2009} using the scaling algorithm developed by \citet{Angulo2010}, which is scaled to WMAP-7 yr from WMAP-1 yr cosmology. The MRIIscWMAP7 follows 2160$^{3}$ dark matter particles in a smaller box with higher resolution. The box is 104.311 $\rm Mpc$/$h$ each side and the particle mass is $\rm m_{p}=8.5024 \times 10^{6}$ $\rm M_{\odot}$/$h$. The MRIIscWMAP7 has a 110 times better mass resolution than the MR7, which is sufficient to study the properties of satellite galaxies around the main member galaxies of the LG, the MW and M31 analogs.

At each simulation snapshot, dark matter halos are grouped using a friend-of-friend (FOF) algorithm \citep{Davis1985}, with a linking length equal to 0.2 times the mean interparticle separation. Then SUBFIND algorithm \citep{Springel2001} is applied to each FOF group to find a self-bounded substructure (subhalo). In each FOF group, the galaxy hosted by the main subhalo is referred to as the central galaxy, and the others are referred to as satellite galaxies.

Baryonic processes have a significant impact on the galaxy formation and properties of a galaxy. The semi-analytical galaxy formation model of \citet{Guo2013} applied to the halo merger trees extracted from these two $\emph{N}$-body simulations follows the baryonic processes from the very beginning of the simulation to the present day. Most observed properties of galaxy have been reproduced by this semi-analytical galaxy formation model, such as luminosity function, size, clustering, and etc. More detailed information about the semi-analytical galaxy formation model are described in \citet{Guo2011} and \citet{Guo2013}.

For comparison, we also consider a hydrodynamical simulation TNG100-1, which is the central volume simulation of the IllustrisTNG project and includes various baryonic processes in addition to the gravity. 
The side length of TNG100 is 110.7~Mpc with $\rm m_{DM}=7.5 \times 10^{6}$ $\rm M_{\odot}$. It has been proven successful in reproducing many galaxy observables including size, morphology, luminosity function, and etc. Details of the simulation can be found in \citet{Nelson2019}.

\subsection{The LG candidates}

As already mentioned in the introduction, in this work we adopt stellar mass, instead of the usually used halo mass, to identify the MW and M31 analogs. LG candidates (MW-M31 pairs) are then identified after constraining further the separation and kinematics of the MW and M31. 

We adopt the following criteria on stellar masses of the MW and M31:
\begin{equation}
   \rm  4\times10^{10}M_{\odot}<M_{\rm MW}<8\times10^{10}M_{\odot} \, ({\rm MW});
	\label{eq1}
\end{equation}
\begin{equation}
    \rm 8\times10^{10}M_{\odot}<M_{\rm M31}<13\times10^{10}M_{\odot} \, ({\rm M31}).
	\label{eq2}
\end{equation}
The lower and upper bounds of stellar masses of the MW and M31 are set on the basis of the results from \citet{McMillan2011} and \citet{Sick2015}, who determine the stellar masses of the MW and M31 to be 6.4$\times$10$^{10}$M$_{\odot}$ and 10.3$\times$10$^{10}$M$_{\odot}$, respectively.  
There are 382,560 MW and 103,157 M31 analogs in the MR7, among which 247,002 and 71,871 are central galaxies (type=0), as listed in Table~\ref{table1}. In the MRIIscWMAP7 catalog, the corresponding numbers are 4055 and 1342, with 2488 and 880 as central galaxies. For TNG100, there are 1295 MW and 459 M31 analogs, with 813 and 301 as central galaxies. Hereafter we refer to the central galaxies that satisfy the MW and M31 selection criteria as the central MW analogs and the central M31 analogs, respectively.

\begin{deluxetable}{lccc}
\tablewidth{0.48\textwidth}
	\centering
	\tablecaption{Numbers of the MW and M31 Analogs Satisfying the Stellar Mass Criterion in the MR7, MRIIscWMAP7, and TNG100 Catalogs.}
	\tablehead{  & \colhead{MR7} & \colhead{MRIIscWMAP7} & \colhead{TNG100}}
    \startdata
	    MW   & 382,560 & 4055 & 1295\\	
		M31   & 103,157 & 1342 & 459\\
	    MW (Central galaxies)   & 247,002 & 2488 & 813\\	
		M31 (Central galaxies)   & 71,871 & 880 & 301
	\enddata
\label{table1}
\end{deluxetable}

We then select MW-M31 pairs by applying criteria on MW-M31 distance and kinematics. \citet{McConnachie2012} found the distance between the MW and M31 to be $\sim$ 787~kpc. Each MW-M31 pair is thus defined as a system consisting of one MW and one M31 analog with separation between 0.6 and 1 Mpc. The MW and M31 analogs are referred to as the main member galaxies. In each MW-M31 pair at least one main member is a central galaxy.
To mimic the large-scale environment of the LG, we adopt the {\sl isolation} criterion of \citet{Fattahi2016}, which requires that there be no halos more massive than the less massive halo of the main members within 2.5 Mpc. The large-scale influence on galaxy formation, if any, is found not to be strong, if any. Here we adopt the halo mass rather than stellar mass in order to focus on the influence on the halo-scale. The results will not be affected significantly.

The relative radial velocity of the MW and M31 has been measured with high accuracy. For example, \citet{vanderarel2012} measure a relative radial velocity of -109.3$\pm4.4$~km ${\rm s^{-1}}$. 
On the other hand, the tangential velocity between M31 and MW is poorly constrained. Results of different works vary from $\sim10$ km ${\rm s^{-1}}$ to $\sim160$ km ${\rm s^{-1}}$ \citep{Sohn2012,Salomon2016,Marel2019}. Here we consider two possibilities: one with relative tangential velocity far less than the relative radial velocity (Mod1: ${\rm v_{tan} < v_{rad}}$), and the other with relative tangential velocity comparable with the relative radial velocity 
(Mod2: ${\rm v_{tan}}$$\sim$${\rm v_{rad}}$). When selecting our sample using radial velocity criteria, we also correct the relative radial velocity for Hubble expansion. For MR7, $v_{tan}= 17 \pm 17$ km/s for Mod1 (radial mode, \citealt{Sohn2012}), and $v_{tan}= 164.4 \pm 61.8$ km/s for Mod2 (low-ellipticity orbit mode, \citealt{Salomon2016}). The selection criteria for MW-M31 pairs are summarized as follows:
\begin{eqnarray}
&0.6~{\rm Mpc} < {\rm d} < 1~{\rm Mpc}\nonumber \nonumber\\
&-190~{\rm km/s} < {\rm v_{rad}} < -135~{\rm km/s}\nonumber\\
& {\rm v_{tan}} \leq  34~{\rm km/s} &({\rm Mod1})\nonumber\\
&103~{\rm km/s} < {\rm v_{tan}} < 226~{\rm km/s}  & ({\rm Mod2}) \nonumber
\end{eqnarray}

For MRIIscWMAP7, in order to have better statistics, we relax the selection criteria of velocities as follows:
\begin{eqnarray}
&0.6~{\rm Mpc} < {\rm d} < 1~{\rm Mpc}\nonumber \nonumber\\
&-218~{\rm km/s} < {\rm v_{rad}} < -108~{\rm km/s}\nonumber\\
&{\rm v_{tan}} \leq  0.5 {\rm v_{rad}} &({\rm Mod1})\nonumber\\
&0.5 {\rm v_{rad} < v_{tan}} < 1.5 {\rm v_{rad}}  & ({\rm Mod2}) \nonumber
\end{eqnarray}

The numbers of LG analogs associated with different selection criteria are listed in Table~\ref{table2}. There are 7480 MW-M31 pair analogs satisfying the separation criteria, yet only 26 (134) LG candidates satisfy the full selection criteria of Mod1 (Mod2) in MR7. The isolation criteria remove 2/3 of the sample, and the radial velocity further remove 5/6 from the remaining sample. Finally, less than 2\% of the MW-M31 pair analogs are classified as LG analogs, corresponding to a volume density of 10$^{-7}$/(Mpc/$h$)$^3$.

We note that no (only one) LG candidate satisfies the full selection criteria of Mod1 (Mod2) in TNG100. Given that the volume of the TNG100 is about 263 times smaller than that of the MR7, the numbers of LG candidates in TNG100 are consistent with those in MR7. We thus do not use the TNG100 LG analog to analyze the galaxy properties. 

\begin{deluxetable}{cccc}
\tablewidth{0.48\textwidth}
	\centering
	\tablecaption{Numbers of LG analogs with different selection criteria in MR7, MRIIscWMAP7 and TNG100.}
	\tablehead{ \colhead{Selection criteria} & \colhead{MR7} & \colhead{MRIIscWMAP7} & \colhead{TNG100} }
    \startdata
		d & 7480 & 100 & 12\\
        d, isolated & 2218 & 38 & 4\\
        d, isolated, ${\rm v_{rad}}$ & 376 & 15 & 1\\
        d, isolated, ${\rm v_{rad}}$, ${\rm v_{tan}}$$<$${\rm v_{rad}}$  & 26 & 10 & 0\\
        d, isolated, ${\rm v_{rad}}$, ${\rm v_{tan}}$$\sim$${\rm v_{rad}}$  & 134 & 3 & 1
	\enddata
\label{table2}
\end{deluxetable}

\section{RESULTS}
In this section, we investigate the host halo, main member galaxy properties of the LG analogs, and the satellite galaxies around the two main members.

\subsection{Halo mass}
Halo mass is one of the most important properties in understanding the formation of the LG and the MW \citep{Cautun2014,Carlesi2017}. It is usually estimated using kinematics of the satellite galaxies and halo stars \citep[e.g.][]{Wilkinson1999,Sakamoto2003,Xue2008,Wang2015,Zhai2018}. Here we use the semi-analytic model to study the halo mass distribution and its dependence on different properties. In this paper, halo mass is referred to as the virial mass, which is defined as the mass enclosed within R200, where R200 is the radius within which the averaged density is 200 times the critical density.

\begin{deluxetable*}{lcccccc}
\tablewidth{0.9\textwidth}
	\centering
	\tablecaption{Mass Distributions of the LG Analogs and Their Main Members.}
	\tablehead{  &  & \colhead{Mod1} &  &  & \colhead{Mod2} & }
    \startdata
	&	M$_{{\rm M31}}$	&	M$_{{\rm MW}}$	&	M$_{{\rm LG}}$	&	M$_{{\rm M31}}$	&	M$_{{\rm MW}}$	&	M$_{{\rm LG}}$	\\
		\hline
		Median mass	&	2.5	&	1.5	&	4.4	&	3.8	&	2.5	&	6.6	\\
$1-\sigma$ region	&	[1.4, 3.8]	&	[0.8, 2.9]	&	[2.5, 6.3]	&	[2.0, 6.6]	&	[1.1, 4.7]	&	[4.7, 10]	\\
		\hline
		Median mass (M$_{\rm bulge}$/M$_{\ast}<0.5$)	&	2.0	&	1.4	&	3.7	&	3.0	&	1.5	&	6.0	\\
$1-\sigma$ region	&	[1.2, 2.7]	&	[0.8, 3.0]	&	[2.3, 5.7]	&	[1.8, 6.0]	&	[1.1, 4.1]	&	[4.2, 8.5]
    \enddata
    \tablenote{ We list the median halo mass and $1-\sigma$ range. The first two rows are for the MW, M31 and the LG analogs and the second two are for those if further requiring the MW and M31 analogs to be disk-dominated systems. All the masses are in the units of 10$^{12}\rm M_{\odot}$.}
\label{tableM}
\end{deluxetable*}

\subsubsection{Host halo mass of M31 and the MW}
In previous re-simulations of the MW and the M31, they are usually chosen as central galaxies. As members of the LG, they could have different mass distributions.

\begin{figure*}

	\includegraphics[width=\textwidth, trim=40 250 50 260, clip]{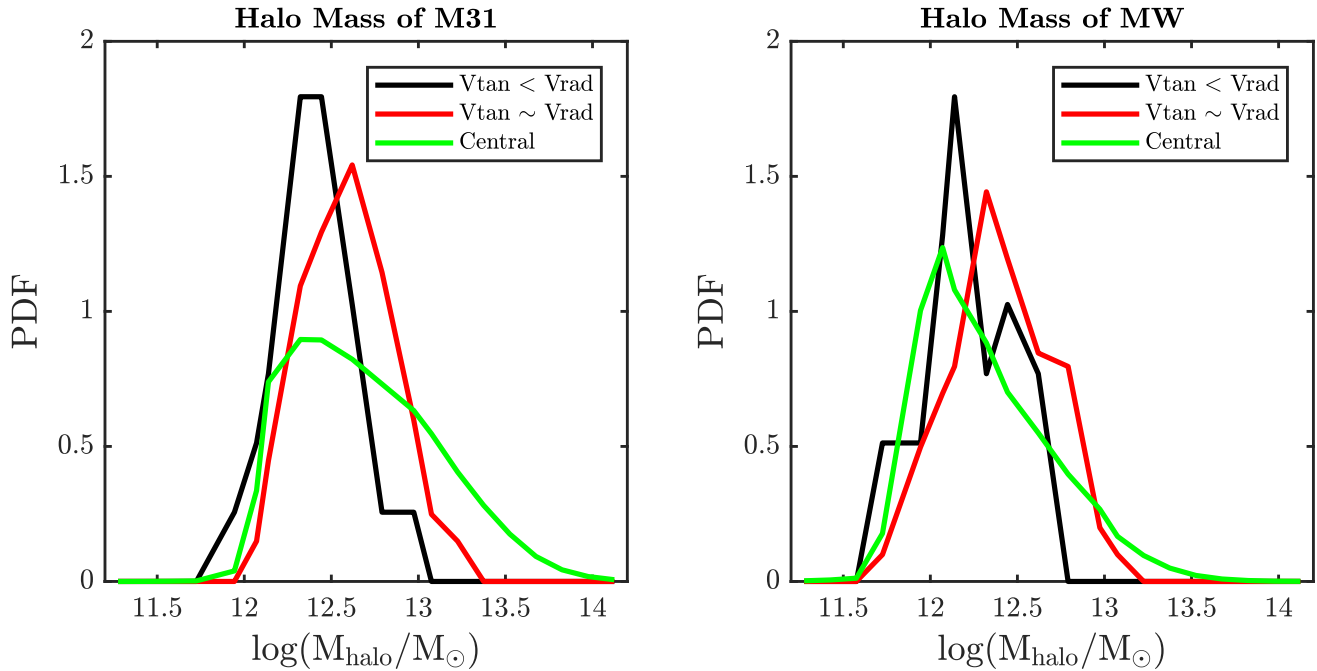}
    \caption{
    Probability distribution functions of halo masses for M31 (left) and the MW (right) analogs in MR7. 
    The black curves are for Mod1, the red curves are for Mod2, and the green curves are for the central MW/M31 analogs.}
    \label{fig1}
\end{figure*}

Fig.~\ref{fig1} shows the probability distribution functions (PDFs) of halo mass for each member of the LG analogs, M31 (left) and MW (right) analogs respectively. For the radial orbits, the median halo mass of M31 is about 2.5$\times$10$^{12} \rm M_{\odot}$, which is consistent with former works \citep{Watkins2010,Fardal2013}, while the median halo mass of the MW is about 1.5$\times$10$^{12} \rm M_{\odot}$, slightly higher than the results in the literature \citep[e.g.][]{Wilkinson1999,Smith2007,Xue2008,Zhai2018,posti2019}. A further constraint on the bulge-mass fraction reduces the corresponding host halo mass slightly (see Table~\ref{tableM}). The halo masses of each LG member in Mod2 are higher than those in Mod1, suggesting that the tangential velocity has influence on the halo mass distribution of each LG member. We perform a two-sample Kolmogorov-Smirnov (K-S) test on the PDFs of halo masses in Mod1 and Mod2, and we find that the null hypothesis, i.e., that the distributions of Mod1 and Mod2 have no difference can be rejected with $p$-values of 0.005 and 0.03 for M31 and the MW, indicating that the difference between the two modes is statistically significant at a confidence level $>$97\%. A two-sample $t$-test also indicates that the mean values of the two distributions are significantly different, with $p$ = 0.001 and 0.005 for M31 and the MW. The peak mass of the central M31/MW analogs is similar to their corresponding members in Mod1 but smaller than those in Mod2. Both central M31 and MW analogs show a wider range of dark matter halo mass, extending to higher masses. The lack of high-mass analogs for the LG members might be due to the restricted isolation criteria.

\begin{figure*}

	\includegraphics[width=\textwidth, trim=40 250 50 260, clip]{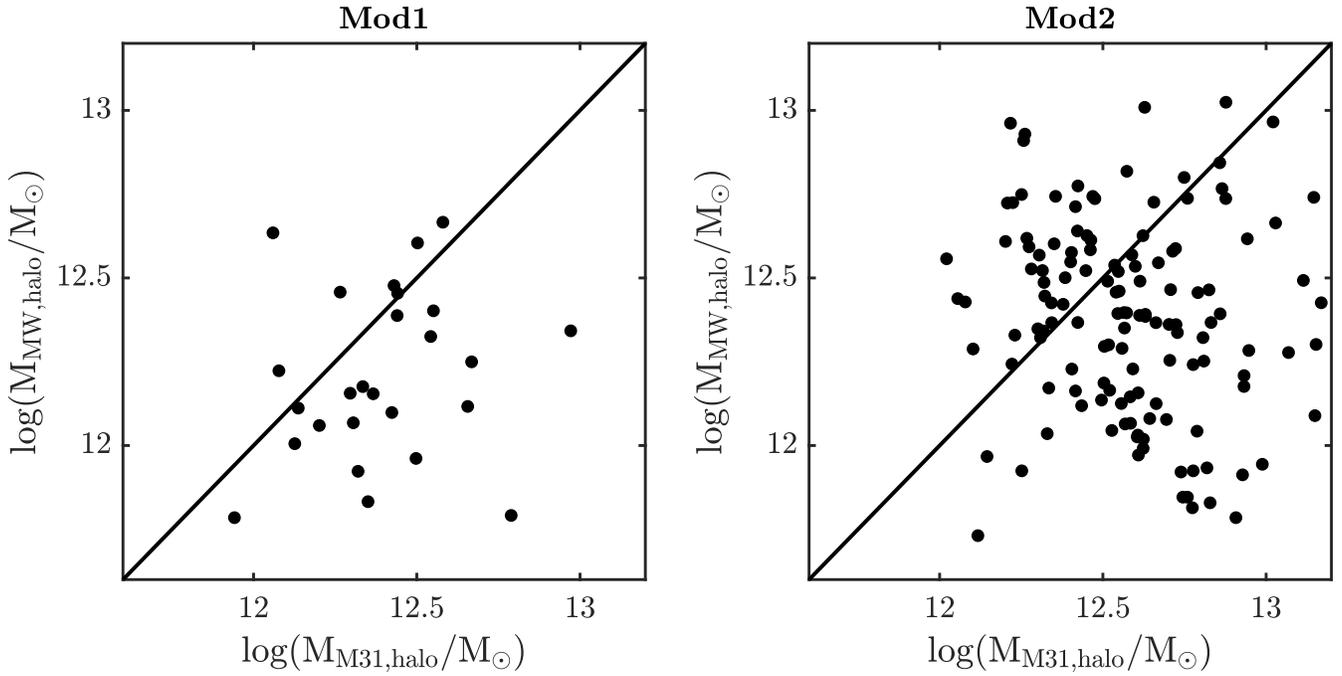}
    \caption{Left panel: scatter plot of the M31 halo mass to the MW halo mass in Mod1. Right panel: the same in Mod2. The black solid line is the bisectrix.}
    \label{figscat}
\end{figure*}

Although M31 has larger stellar mass than the MW, its halo mass is not necessarily higher. We show in Fig.~\ref{figscat} the scatter plots of M31 halo mass vs. MW halo mass as the LG members. There are $\sim$27\% LG pairs in Mod1 in which the halo mass of the M31 is smaller than that of the MW, and this fraction increases to $\sim$34\% in Mod2. The K-S test indicates that the difference between the two distributions are insignificant, with a $p$-value of 0.75. A consistent result is also provided by a $t$-test on the mean value of the $\rm M_{MW,halo}/M_{M31,halo}$ ratio, which returns $p=$0.92. Therefore, we conclude that, in both cases, there is the possibility that the halo mass of M31 is smaller than that of the MW.

\subsubsection{Total mass}

\begin{figure*}
	\includegraphics[width=\textwidth, trim=70 310 90 320, clip]{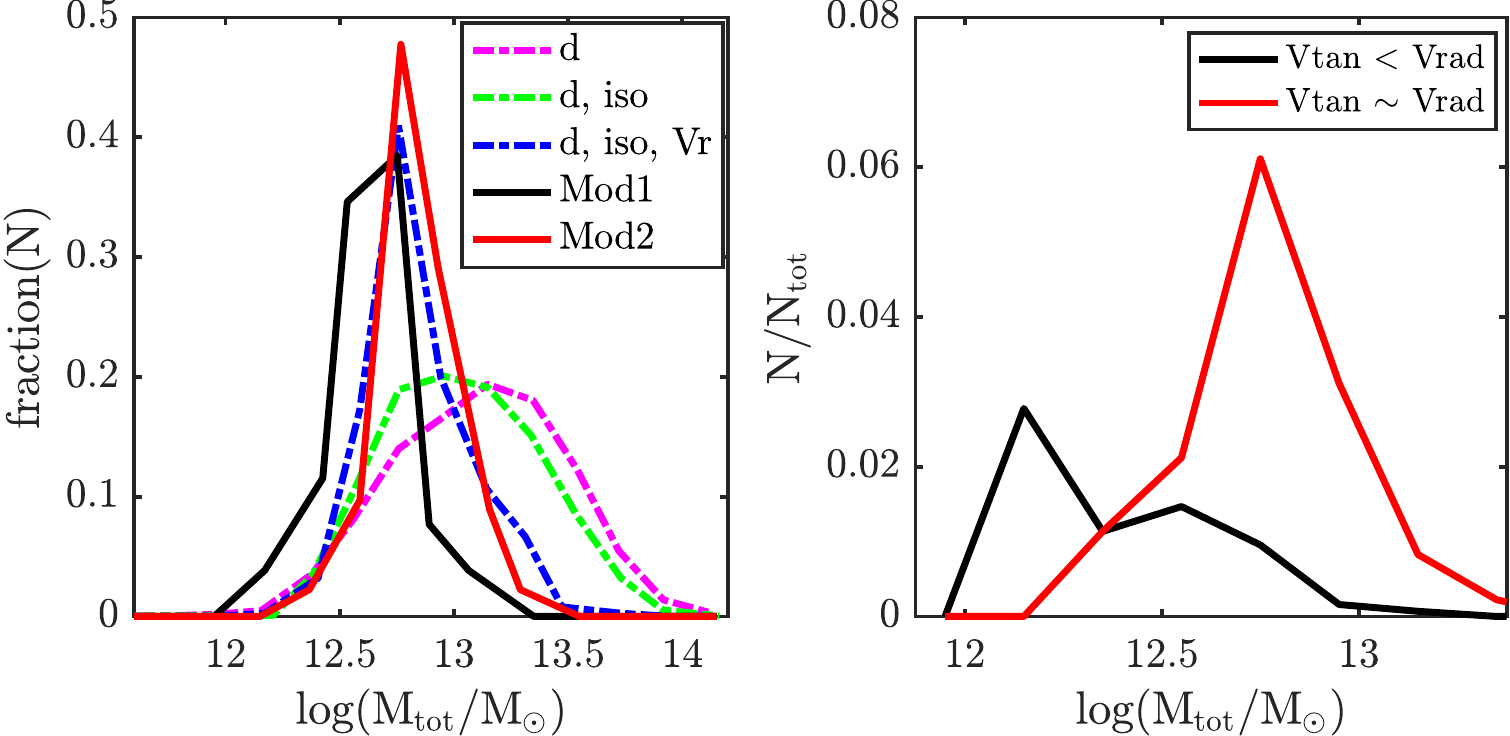}
    \caption{Left panel: total mass distribution of the MW-M31 pairs selected from the MR7. The magenta dashed curve shows the MW-M31 pairs that satisfy the separation criterion. The green dashed curve shows the MW-M31 pairs that satisfy the separation and isolation criteria. Blue dashed curve shows pairs further satisfying the relative radial velocity criterion. The black curve is the total mass distribution of LG analogs satisfying all selection criteria as described in Mod1, while the red curve is for those satisfying criteria in Mod2. 
    Right panel: formation probability of an LG analog (number of LG analogs/number of MW-M31 pairs satisfying the separation criterion) as a function of the total host mass. The black curve is for Mod1, and the red curve is for Mod2.}
    \label{fig3}
\end{figure*}

The left panel of Fig.~\ref{fig3} shows the total mass of the M31-MW pairs under various selection criteria. The total mass is defined as follows. If the two main members of a MW-M31 pair are both central galaxies, the total mass is the sum of the halo masses of the MW and M31 analogs; if one of them is a satellite galaxy, the total mass is defined as the sum of the halo masses of the two galaxies estimated at the time just before their merging. The typical total mass decreases under more and more restricted criteria. The isolation criterion reduces the typical halo mass from 1.3$\times$10$^{13} \rm M_{\odot}$ to 8$\times$10$^{12} \rm M_{\odot}$. Under this isolation criterion, the typical total mass estimated in TNG100 is about 6.5$\times$10$^{12} \rm M_{\odot}$, which is less massive than those in MR7.
The restriction on radial velocity helps further to reduce the total masses, peaking at 5.6$\times$10$^{12} \rm M_{\odot}$. The black curve represents the total mass distribution of the LG analogs with radial orbits, while the red curve represents those with larger tangential velocity. The former tends to be less massive than the latter. The median halo mass of the LG analogs is 4.4$\times$10$^{12} \rm M_{\odot}$ in Mod1 and 6.6$\times$10$^{12} \rm M_{\odot}$ in Mod2. The K-S test indicates that the difference between the two distributions has a high significance, with $p<0.001$. Additionally, according to a $t$-test, the mean values of the LG halo mass for Mod1 and Mod2 are also different at a high confidence level, with $p<$0.001. This result can be interpreted immediately that, given a fixed radial velocity, a higher tangential velocity implies a higher total velocity, which then leads to a higher halo mass. This is consistent with the conclusion of \citet{Carlesi2017}.  

The right panel of Fig.~\ref{fig3} shows the formation probability of an LG analog for a given total mass. Here N$_{\rm tot}$ is the number of MW-M31 pairs satisfying the separation criteria in a given total mass range. The maximum probability of hosting an LG analog is only about 3.0\% at 1.5$\times$10$^{12} \rm M_{\odot}$ for Mod1, and about 6.1\% at 5.6$\times$10$^{12} \rm M_{\odot}$ for Mod2. The M31-MW pairs are rejected as LG analogs predominately by the isolation criterion and then further by the radial velocity criterion.

We summarize the median halo masses and their $1-\sigma$ regions of the LG, M31 and MW analogs in Table~\ref{tableM}.

\subsection{Bulge fraction}
\begin{figure*}
	\includegraphics[width=\textwidth, trim=40 250 50 260, clip]{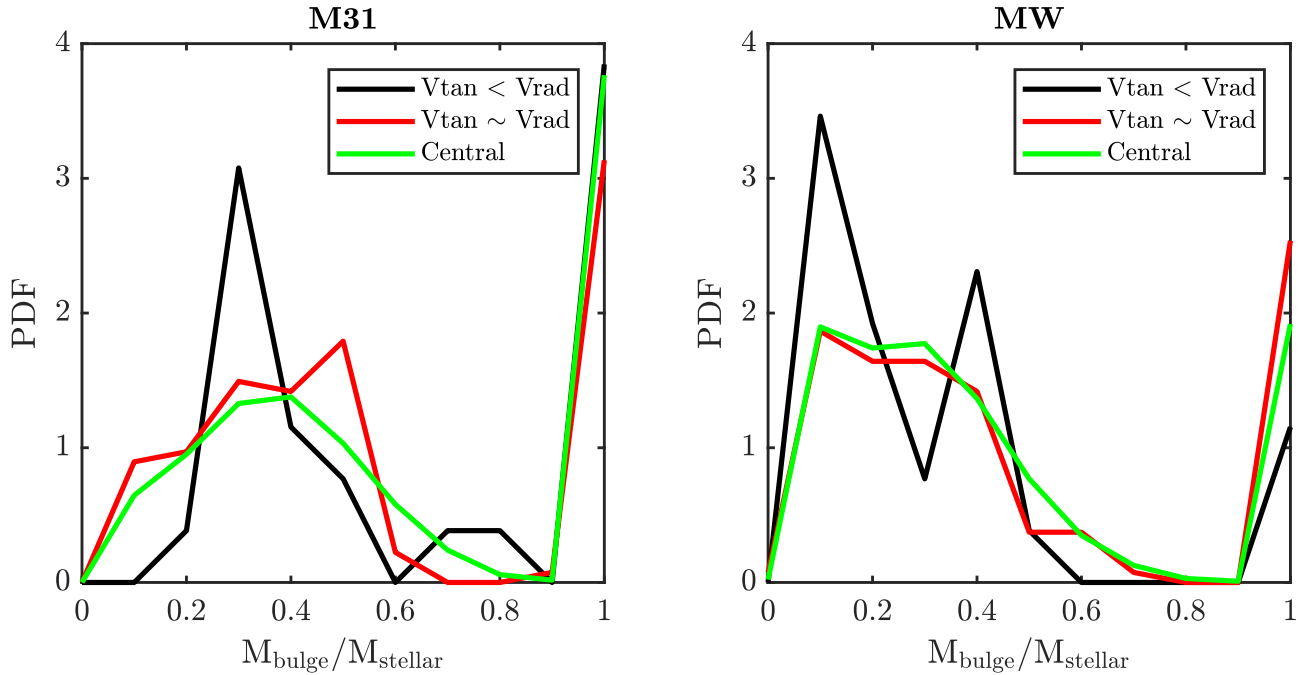}
    \caption{Probability distribution function for the bulge-to-stellar mass ratio of the M31 (left panel) and MW (right panel) analogs. The black curve is for Mod1, the red curve is for Mod2, and the green curve is for the central M31 (Eq.~\ref{eq1}) and MW analogs (Eq.~\ref{eq2}).}
    \label{fig4}
\end{figure*}

We show in Fig.~\ref{fig4} the ratio of bulge stellar mass to total stellar mass (M$_{\rm bulge}$/M$_{\ast}$, bulge fraction hereafter) for each member MW-M31 analog. We also include the central MW-M31 analogs for comparison.

The PDFs of bulge fraction are very similar among these three samples for both the MW and M31. For the MW, the peak of the bulge fraction is about 0.2, and the majority ($>75\%$) of MW analogs are disK-dominated (M$_{\rm bulge}$/M$_{\ast}<0.5$). For M31 (left panel), the peak of the bulge fraction is higher, $\sim$ 0.4, and about half of them are disK-dominated. Interestingly, in Mod1 the distribution of the bulge fraction is rather narrow for the M31 analogs compared to the other mode and the central M31 analogs.

Note that, there is a non-negligible fraction of the MW and M31 analogs as elliptical galaxies in our LG candidates with M$_{\rm bulge}$/M$_{\ast}\sim$ 1. Yet the M31 and the MW are both observed to be disc-dominated. We find the possibility for the LG analogs to host two disk-dominated main members is 50\% for Mod1, while the fraction for Mod2 is 46\%.

\subsection{The anisotropy parameter}
The velocity anisotropy parameter $\beta$ is an important parameter when deriving the halo mass using the Jeans equation.
It is thus crucial to explore the $\beta$ profile of the MW and of the M31, and their dependence on orbital properties. Following the definition in literature \citep{Binney2008}, we define $\beta$ as

\begin{equation}
    \beta=1-\frac{\sigma_{\theta}^2+\sigma_{\phi}^2}{2\sigma_{r}^2}
	\label{eq3}
\end{equation}

This parameter quantities the degree of radial anisotropy in spherical coordinates, where $\sigma_{r}$, $\sigma_{\theta}$ and  $\sigma_{\phi}$ are the velocity dispersions in the radial and two tangential directions. Circular orbits ($\sigma_{r}$=0) correspond to $\beta$=-$\infty$; radial orbits ($\sigma_{\theta}$=$\sigma_{\phi}$=0) correspond to $\beta$=1; isotropic velocity distributions correspond to $\beta$=0.

\begin{figure}
	\includegraphics[width=\columnwidth, trim=80 200 90 200, clip]{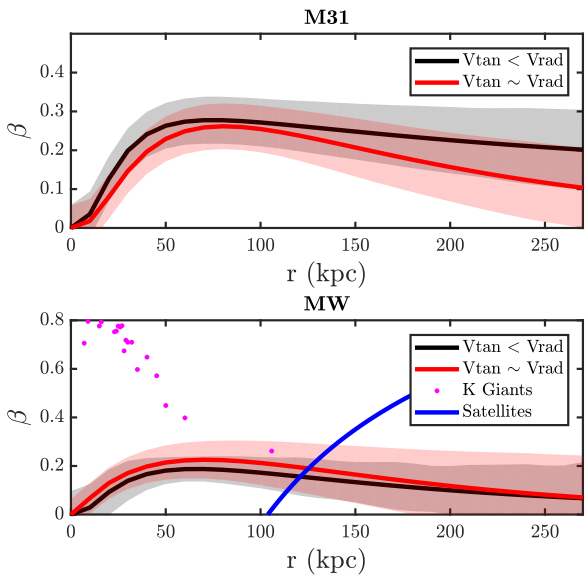}
    \caption{Velocity anisotropy parameters as a function of radius for M31 analogs (upper panel) and the MW analogs (lower panel). The black and red curves denote the $\beta$ profile of dark matter particles in Mod1 and Mod2, respectively. The shaded area denotes the scatter of the profile.} The magenta dots in the lower panel represent the $\beta$ profile computed from the halo K giants from LAMOST \citep{Bird2018}. The blue curve represents the $\beta$ profile of satellite galaxies in $Gaia$ Data Release 2 \citep{Riley2018}.
    \label{figbeta}
\end{figure}

We calculate the velocity anisotropy parameter for the halo using dark matter particles in MR7 (see 
Fig.~\ref{figbeta}). The mean $\beta$ increases with radii and reaches a maximum value of $\sim$0.25 at $\sim$80~kpc for M31, and $\sim$0.2 at $\sim$50~kpc for the MW. It then decreases slowly toward larger radii. Dark matter particles exhibit a rather isotropic radial profile of $\beta$, consistent with results in \citet{Abadi2006}, \citet{Sales2007}, and \citet{Wang2015}. 

The isotropy of the orbit does not affect that of the dark matter velocity field for both M31 and the MW. In the outskirts of M31, $\beta$ differs by a factor of 2 between the two modes of orbit (see the top panel of Fig.~\ref{figbeta}). However, the difference is well within the large scatter.

In the lower panel of Fig.~\ref{figbeta} we show that the $\beta$ profile of the MW differs between different tracers.  Unlike the rather isotropic velocity distribution traced by dark matter, the $\beta$ profile traced by the halo K giants revealed by the LAMOST \citep{Bird2018} showed that their orbits are radial at the center and are nearly isotropic at larger radii. The anisotropy
remains almost constant from approximately 5 to 25 kpc (magenta dots in Fig.~\ref{figbeta}). \cite{Riley2018} used proper motions of 38 satellite galaxies from $Gaia$ Data Release 2 and found that the satellites have tangentially biased motions ($\beta \sim -2$ at 20~kpc) at small radii and radially biased motions ($\beta \sim 0.5$ at 200~kpc) at larger radii (blue line in Fig.~\ref{figbeta}). One explanation for the variance in $\beta$ profile computed with different tracers is the different origin of the tracers: halo stars originate from accreted subhalos; satellite galaxies close to the MW might have been stripped/disrupted; and dark matter particles are from both clumpy and smooth accretion. Interestingly, values of $\beta$ estimated from K giants, satellite galaxies, and dark matter particles agree at $\sim$0.2 at around 120 kpc.

\subsection{Large-scale environments}
We also investigate the large-scale environments of the LGs with the characteristic environmental overdensity: 
\begin{equation}
    \delta=\frac{\rho}{\bar{\rho}}-1,
	\label{eq7}
\end{equation}
where $\bar{\rho}$ is the average matter density over the whole volume $L^3 = (500~{\rm Mpc}/h)^3$. Around each object, $\rho$ is estimated averaged over its surrounding cell of (500/128)~Mpc$/h$ each side.

\begin{figure}

	\includegraphics[width=\columnwidth, trim=100 240 120 250, clip]{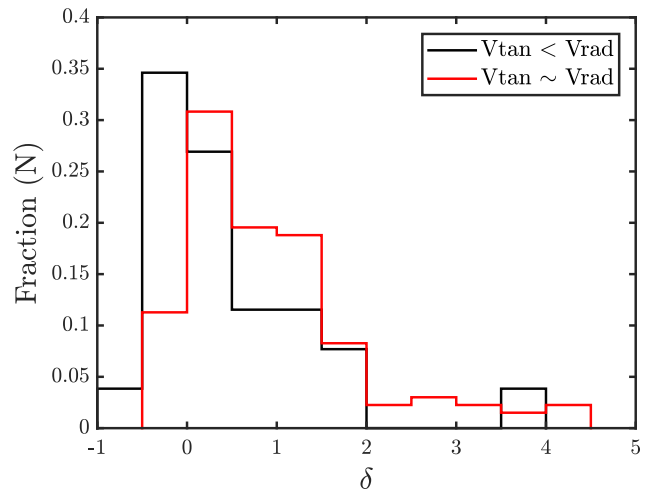}
    \caption{Overdensity distribution of the LG analogs averaged over 4~Mpc. The black histogram is the result of the 26 LG analogs in Mod1, and the red curve is the result of the 134 LG analogs in Mod2.}
    \label{fig6}
\end{figure}

The distribution of environmental overdensity where LGs are formed is shown in Fig.~\ref{fig6}. Black and red histograms are the results for the radial and low-ellipticity orbits, respectively.
Most of the LG analogs are formed at $-0.5<\delta<2$, close to the range of filaments, sheets, and walls \citep{Yan2013,Cautun2014b}. The overdensity of the LGs in radial orbit mode tends to peak at lower values than in the low-ellipticity orbit mode, with a peak $\delta$ of $\sim$0 for the former, and $\sim$0.5 for the latter. We further quantify the difference in the two distributions with the fraction of $\delta<0$, and find the fraction is 38\% for radial orbit mode, and only 11\% for low-ellipticity orbit mode. Given the number of LG analogs in the two modes (26 and 134), the null hypothesis that the two distributions are consistent is rejected with $p\sim0.02$ by the K-S test, and $p\sim0.04$ by the $t$-test, indicating that the difference is significant at a $>2\sigma$ confidence level. 

\subsection{Properties of satellite galaxies}
The mass resolution of MR7 is not sufficient to resolve the satellites. In contrast, the mass resolution of the MRIIscWMAP7 is 110 times better than that of MR7 and can thus be used to study the properties of the satellite populations in the LG. There are in total 10 LG candidates for Mod1, and 3 LG candidates for Mod2.

\begin{figure*}

	\includegraphics[width=\textwidth, trim=40 280 40 260, clip]{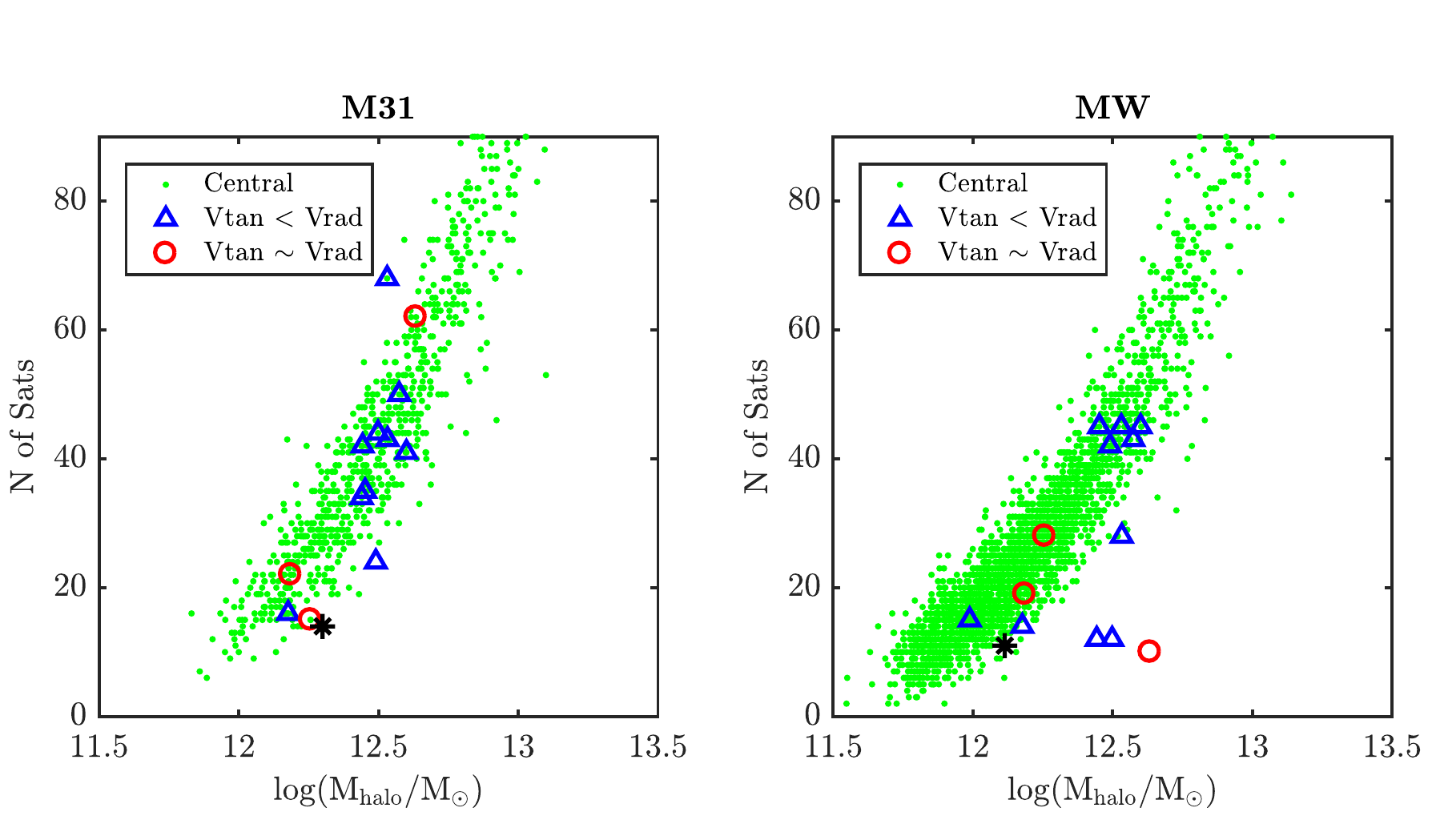}
    \caption{Relation between the number of satellite galaxies and their host halo mass. Left panel and right panel are for M31 analogs and the MW analogs, respectively. Blue and red circles are numbers of satellites in Mod1 and Mod2. Results for satellite galaxies around the central MW/M31 analogs are shown with green dots. Black asterisks denote the observational results: $\rm M_{halo}=2.0\times10^{12}M_{\odot}$ \citep{Fardal2013} and N=14 \citep{McConnachie2012} for M31; $\rm M_{halo}=1.3\times10^{12}M_{\odot}$ \citep{posti2019} and N=11 \citep{McConnachie2012} for the MW.}
    \label{fig7}
\end{figure*}

We select galaxies that lie within 300~kpc from each member galaxy and define them as the corresponding  satellite galaxies. All types of satellite galaxies are selected, including the orphan galaxies, but only those with $\it{M_V}<$ -9 are considered in order to match to the completed sample from observations. Fig.~\ref{fig7} shows that the number of satellites increases rapidly with their host halo mass. This confirms the previous studies where, within less massive halos, the `missing satellite' problem and the `too-big-to-fail' problem could be alleviated. 

\subsubsection{Luminosity function}
\begin{figure*}

	\includegraphics[width=\textwidth, trim=40 240 40 230, clip]{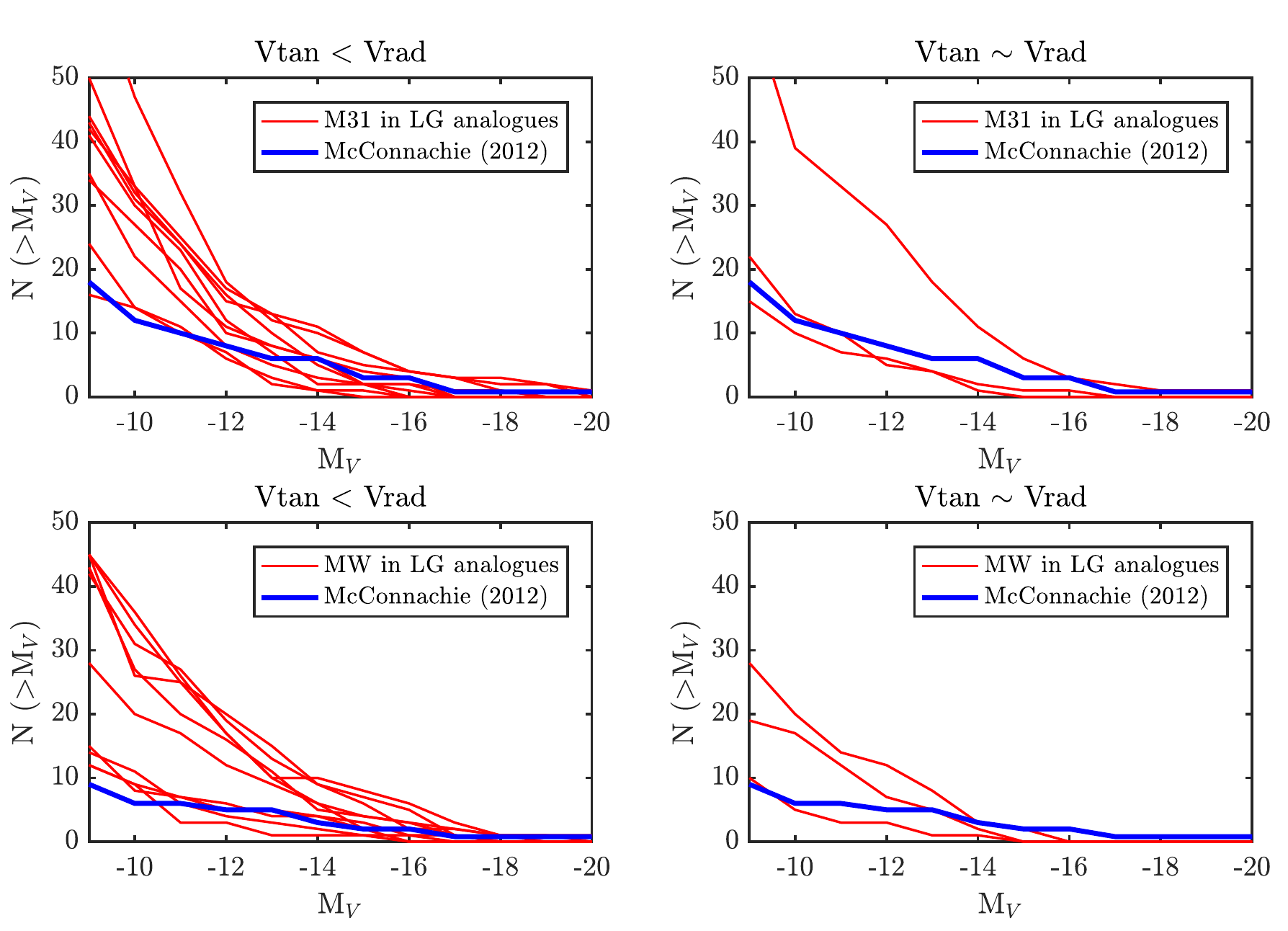}
    \caption{Cumulative number of satellite galaxies vs. $V$-band magnitude for M31 (upper panels) and the MW (bottom panels) as members in the LG analogs (red). The blue solid curve is the observational result from \citet{McConnachie2012}.}
    \label{fig8}
\end{figure*}

We show the $V$-band cumulative number of galaxies as a function of luminosity for satellite galaxies around the M31 analogs (upper panels) and the MW analogs (bottom panels) in Fig.~\ref{fig8}. We make a comparison between the luminosity functions from our simulations and that from the observations of \citet{McConnachie2012}.
In general, the simulated luminosity functions at high luminosity ($M_V < -12$) coincide well with observations for M31 and the MW, both in Mod1 and in Mod2. On the other hand, our simulations predict a larger number of satellites than the observed result at lower luminosity. 

\begin{figure}

	\includegraphics[width=\columnwidth, trim=110 260 120 240, clip]{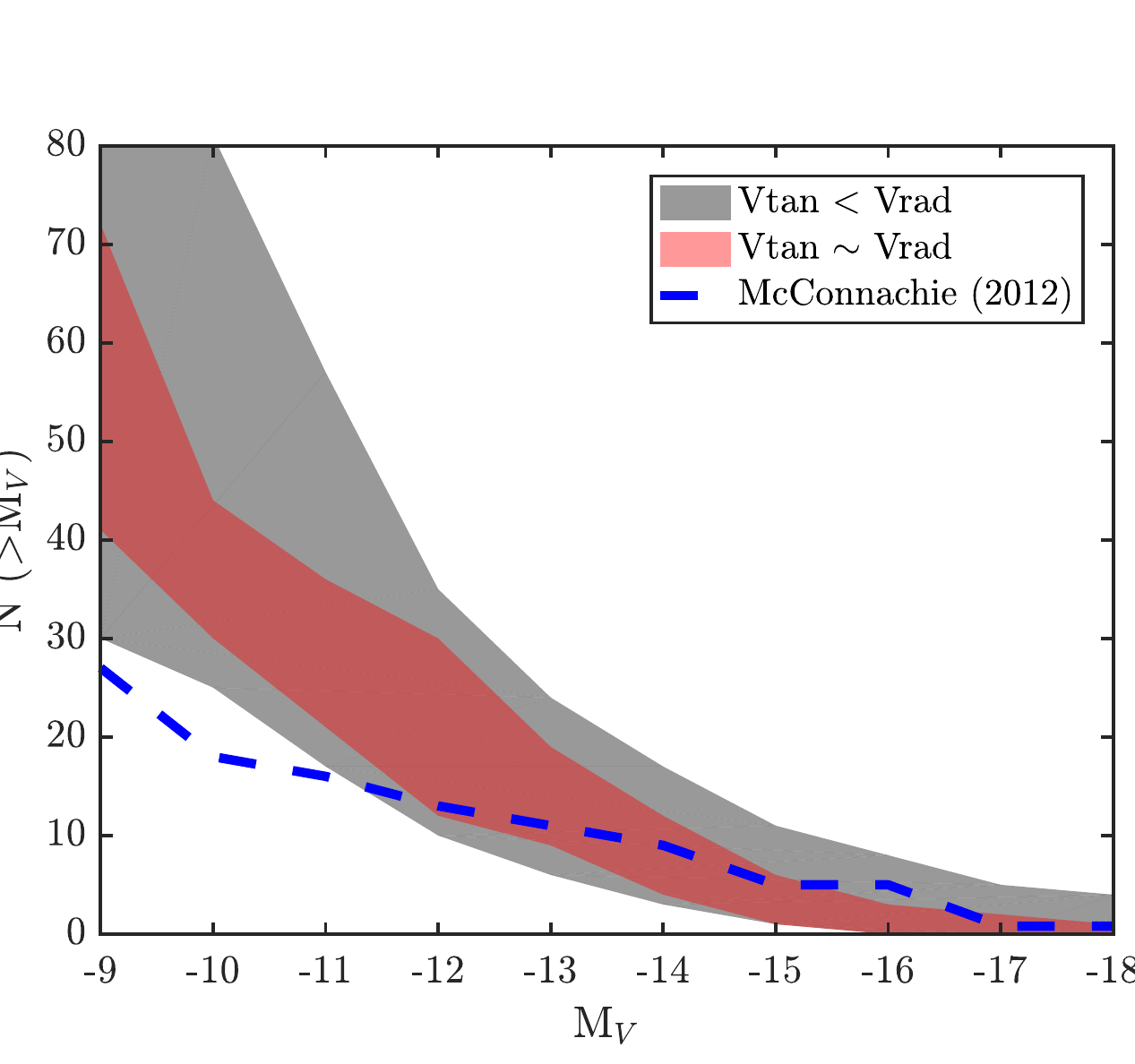}
        \caption{Cumulative number of satellite galaxies vs. $V$-band magnitude for all LG satellites. The black and red shaded areas enclose satellite profiles of all the LG analogs in Mod1 and in Mod2, respectively. The blue dashed curve represents the observational result from \citet{McConnachie2012}.}
    \label{fig9}
\end{figure}

We also compute the $V$-band luminosity function for all the LG satellites. For each mode, we enclose the satellite luminosity function for each LG analog within the corresponding shaded area in Fig.~\ref{fig9}. We find that the amplitude of the cumulative distribution profile of $V$-band magnitude is slightly lower for Mod2 compared to Mod1. Similar to Fig.~\ref{fig8}, we also find that both the two modes show an enhancement at low luminosity with respect to the observational results of \citet{McConnachie2012}.

\subsubsection{Radial profile}

\begin{figure*}
	\includegraphics[width=\textwidth, trim=40 240 40 240, clip]{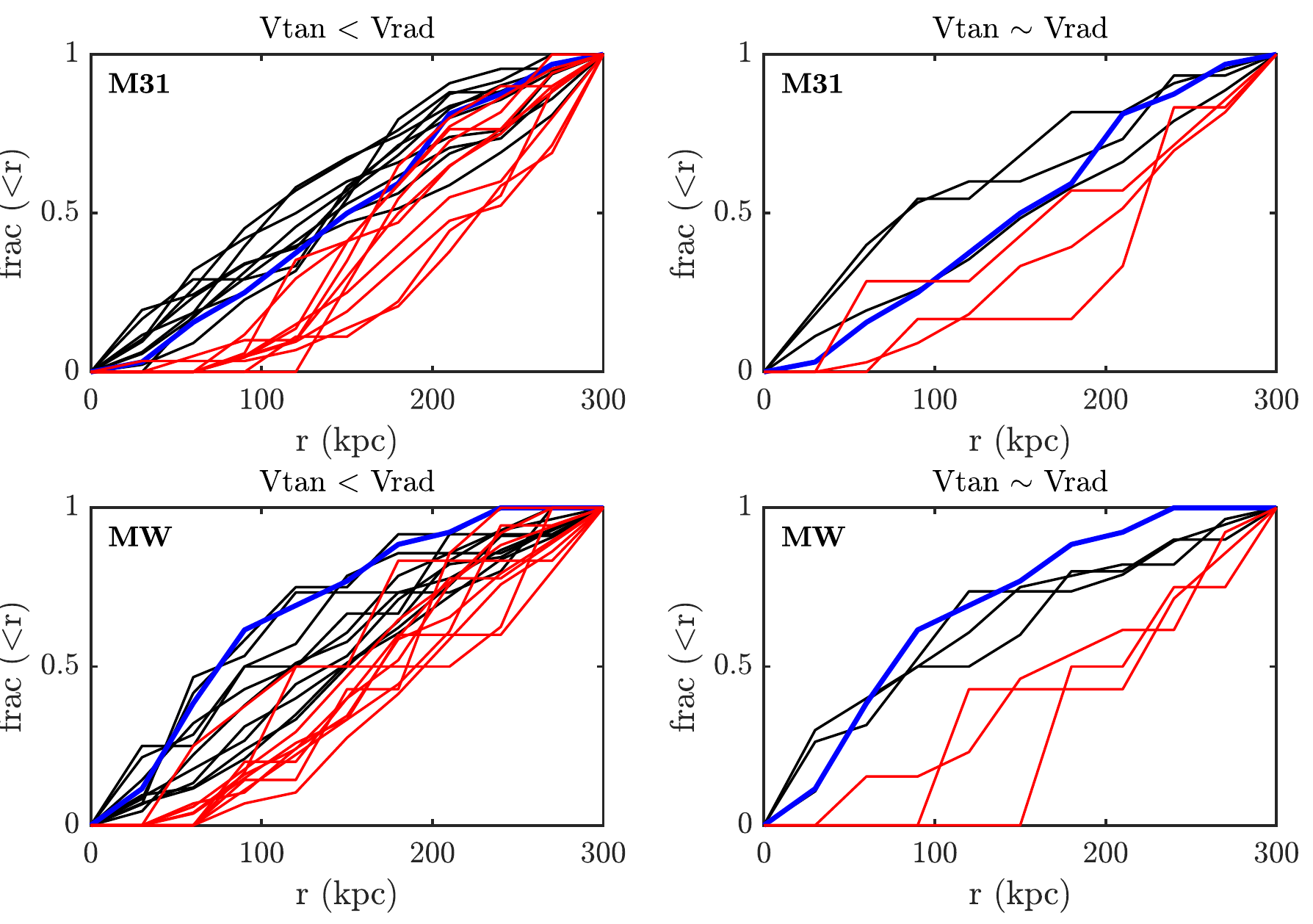}
    \caption{Cumulative radial profiles of the M31 satellites (upper panels) and the MW satellites (bottom panels) as members of the LG analogs in the MRIIscWMAP7. The left panels are for the Mod1 and the right panels are for the Mod2. Black and red curves denote the profiles with and without orphan galaxies. The blue solid curves show the observational results from \citet{McConnachie2012}. }
    \label{fig10}
\end{figure*}

Besides the luminosity function, the spatial distribution of satellite galaxies around the LG is also a key point to verify the $\Lambda$CDM model. The positions of orphan satellite galaxies are unreliable \citep{2017Pawlowski} if they are simply tied to single dark matter particles. In \cite{Guo2011} they assumed a constant mass satellite spirals to the center of the host halo with the `isothermal’ profile on a near-circular orbit. It was predicted that the distance to the host center decays with the square root of the time. In practice, the distance of the orphan galaxy to the host center is the positional offset of the tracer particle from the central galaxy multiplied  a factor of $\sqrt{1-t/t_{\rm friction}}$ , where {\sl t} is the time since the merger clock was initialized and $t_{\rm friction}$ is the dynamical friction time. To test the influence of the orphan galaxies, here we also consider two satellite samples, with and without orphan galaxies, when calculating the spatial distribution of satellite galaxies.

In Fig.~\ref{fig10} we compare the cumulative radial distribution of predicted satellite galaxies with $M_V<$ -9 with the observed ones for the M31 (upper panels) and the MW analogs (lower panels). The observational results are from \citet{McConnachie2012}, shown as blue solid lines. In both modes, the model predictions agree broadly with the observational results, and it is difficult to tell which mode is more preferred to reproduce the radial distribution of satellite galaxies around the MW and M31. The difference between distributions with and without orphan galaxies in Fig.~\ref{fig10} shows 
the distribution exhibits a central peak if the orphan galaxies are included. 
In \cite{Guo2011}, they already showed the satellite profile converges in clusters between the MS and the MS-II, which suggests the simple treatment of the orphan galaxies works reasonably well.

However, we recall that a caveat here is the influence of the central galaxy's potential on the distribution of satellites, which is particularly important for low-mass satellite galaxies as considered here. The central galaxy's potential may destroy the satellites in the inner regions, and reduce the number of satellites by $\sim$40\% within 50 kpc, although the fraction decreases rapidly in outer regions \citep{Garrison2017,swala2017,Kelley2019}. This would imply that the central peak in the radial distribution of  satellites could be even shallower than those shown in Fig.~\ref{fig10}. 

\section{DISCUSSION}

LG analogs studied in simulations are usually defined as halo pairs satisfying a certain range of the halo mass. 
\citet{Carlesi2017} calculate the halo masses of M31 and MW, and constrain the total mass of the LG to be $2.57_{-1.19}^{+0.87}\times 10^{12} \rm M_{\odot}$. Consistently, \citet{Gonz2014} and \citet{Pen2014} conclude that the LG favors a total mass of $2.40_{-1.05}^{+1.95}\times 10^{12} \rm M_{\odot}$ and $2.3_{-0.7}^{+0.7}\times 10^{12} \rm M_{\odot}$. \citet{Li2008} measure the calibrated-timing-arguments-estimated LG mass to be log(M$_{\rm LG}/\rm M_{\odot}) = 12.27^{+0.31}_{-0.44}$.
The total mass of LG selected with stellar mass in our work is $4.4_{-1.9}^{+1.9}\times 10^{12} \rm M_{\odot}$, or {\rm log}(M$_{\rm LG}/\rm M_{\odot}) = 12.64_{-0.24}^{+0.16}$, slightly larger than the previous results, although all the results are broadly consistent considering the large error bars.

The mass of the MW obtained using simulations is very close to or slightly lower than $10^{12}\rm M_{\odot}$ \citep{Pen2014,Patel2017,Carlesi2017,Patel2018,Callingham2018}. In our work, we get a higher value: $1.5_{-0.7}^{+1.4} \times10^{12}\rm M_{\odot}$. On the other hand, the mass of M31 we find in our work is consistent with previous studies \citep{Fardal2013,Pen2014,Patel2017,Carlesi2017}.

It is clear to see from Fig.~\ref{fig7} that the number of bright satellites ($\rm M_V < -9$) strongly correlates with halo mass. This result is consistent with \citet{Fattahi2016}, who used the total halo mass of MW-M31 pairs to select LG candidates and obtained the same results. 
The radial distributions of satellites are also in general consistent with those of \citet{Newton2018}, where the simulated satellite distribution matches well with observations from the Sloan Digital Sky Survey (SDSS) and Dark Energy Survey (DES).

\section{CONCLUSIONS}
We use semi-analytic galaxy catalogs based on the MR7 and the MRIIscWMAP7 simulations to study the properties of LG host halos and their satellite populations, for two types of LG analogs distinguished on the basis of the relative orbits between the MW and M31. Our conclusions are summarized as follows.

\begin{itemize}
\item Our results show that the relative tangential velocity between the MW and M31 has great influence on both the total mass of the LG analog and the mass of each member M31 and MW analog. In radial orbits where $\rm v_{tan} < v_{rad}$, the typical host halo mass of the LG, the MW, and M31 are $4.4_{-1.5}^{+2.4}\times$10$^{12}\rm M_{\odot}$, $1.5_{-0.7}^{+1.4}\times$10$^{12}\rm M_{\odot}$, and  $2.5_{-1.4}^{+1.2}\times$10$^{12}\rm M_{\odot}$. In low-ellipticity orbits where $\rm v_{tan} \sim v_{rad}$, the corresponding masses are $6.6_{-1.5}^{+2.7}\times$10$^{12} \rm M_{\odot}$, $2.5_{-1.1}^{+1.3}\times$10$^{12} \rm M_{\odot}$, and $3.8_{-1.8}^{+2.8}\times$10$^{12}\rm M_{\odot}$, respectively. These are generally consistent with recent observations, but slightly higher than previous studies using halo properties to select the LG analogs. 

\item For a given total mass, the probability of hosting LG analogs is very low, with a maximum value of $\sim$6\%. This result indicates the large variance in the property of galaxies with a certain halo mass, and further implies that halo mass alone is not sufficient to identify LG analogs; other quantities, e.g. stellar mass, should also be considered in such studies.

\item The distribution of bulge fraction covers a wide range. The MW analogs tend to be more disk-dominated compared to the M31 analogs. Both M31 and MW analogs have a sub-population with larger bulge-to-total stellar fraction (M$_{\rm bulge}$/M$_{\ast} >$ 0.8). The possibility that the two main member galaxies in the LG analogs are both disk-dominated galaxies is 50\% for Mod1 and 46\% for Mod2.

\item The anisotropic parameter $\beta$ of dark matter particles rises with radius, corresponding to nearly isotropic near the center, reaching $\beta$=0.2--0.3 at $r$= 50--80~kpc, and then decreases slowly toward larger radii. The profile of $\beta$ obtained in our work is distinct from that computed with other tracers, such as halo K giants, and satellites, which reflects the variance in the origin and evolution of different tracers. 

\item Most LG analogs are formed in filaments, with a tail toward higher densities. In particular, systems with larger tangential velocities tend to form in slightly denser regions than those with radial orbits.

\item We confirm previous findings that the total number of satellites increases with host halo mass. Broad agreement is found between observational results and our simulation on the $V-$band luminosity function and radial profiles of the LG's satellite systems. More detailed analysis of satellite properties will require higher-resolution simulations in the future.
\end{itemize}

\acknowledgments

We thank the anonymous referee whose constructive comments substantially improved the paper. We thank Xiangxiang Xue, Chengdong Li, and Jie Zheng for great help and suggestions for this study. 
This work is supported by the National Natural Science Foundation of China under grants Nos. 11988101, 11890694, 11622325, 11573033, 11573032, 11873052, 11973049, 11390371, and the National Key R\&D Program of China No.2019YFA0405502, No.2018YFA0404503 and No. 2017YFB0203300. This work is supported in part by the NAOC Nebula Talents Program.
This work is supported by the Strategic Priority Research Program of Chinese Academy of Sciences, Grant No.XDA15016200.
This work is sponsored (in part) by the Chinese Academy of Sciences (CAS), through a grant to the CAS South America Center for Astronomy (CASSACA) in Santiago, Chile. 
Q.G. acknowledges support from the Royal Society Newton Advanced Fellowships. 
The Millennium Simulation databases used in this paper and the web application providing online access to them were constructed as part of the activities of the German Astrophysical Virtual Observatory (GAVO).

\bibliography{LG}

\end{document}